\documentclass[12pt,a4paper]{article}

\usepackage[T1]{fontenc}
\usepackage{newtxtext,newtxmath}
\usepackage[margin=2.5cm]{geometry}
\usepackage{setspace}
\usepackage{parskip}

\usepackage{graphicx}
\usepackage{amsmath}
\usepackage{booktabs}
\usepackage{caption}
\usepackage{subcaption}
\usepackage{float}
\usepackage{placeins}

\usepackage{tabularx}
\usepackage{array}
\usepackage{makecell}
\usepackage{threeparttable}
\usepackage{ragged2e}

\usepackage[table]{xcolor}
\usepackage{needspace}

\usepackage{tikz}
\usetikzlibrary{
    arrows.meta,
    positioning,
    shapes.geometric,
    calc,
    matrix
}

\usepackage{url}
\usepackage[hidelinks]{hyperref}

\usepackage{titlesec}

\titleformat{\paragraph}[block]
  {\bfseries}
  {}
  {0pt}
  {}

\titlespacing*{\paragraph}
  {0pt}
  {1em}
  {0.6em}

\usepackage[nottoc]{tocbibind}
\usepackage[title]{appendix}

\begin{document}


\begin{center}

{\LARGE\bfseries What Makes Creation Human?}

\vspace{0.3em}

{\large Authorship, Reasons, and Meaningful Human Control in Generative AI}

\vspace{0.6em}

{\small Yuxi Cao\footnote{\texttt{yuxi.cao@psl.eu}}}

\end{center}

\begin{abstract}
Generative artificial intelligence (GenAI) significantly expands creators' productive capacity, but this expansion does not necessarily entail a corresponding increase in creative agency or authorship. Starting from this divergence, this paper distinguishes creativity at the level of the work from creative agency at the level of the creator, and examines how GenAI restructures creative relations at the individual, collective, and industrial levels. It argues that human authorship cannot be determined solely by the amount of manual intervention, the degree of automation, the origin of an initial idea, or the possession of final selection authority. Rather, authorship depends on whether human judgment and reasons genuinely shape the development of the work.

To articulate this requirement, the paper introduces Meaningful Human Control (MHC) into the context of generative creation and identifies a distinctive limitation of its classical tracking condition. In creative processes, authorship is not primarily a static relation between a person and a finished work. A creator's reasons are not always fully specified prior to interaction; they may instead emerge, be revised, or be abandoned through ongoing engagement with AI, while standing in a diachronic normative relation to the trajectory through which the work develops.

To address this problem, the paper proposes dynamic-reflexive tracking (DRT). DRT requires that a creator's reasons be capable of forming and changing over time, undergoing reflective uptake, and exerting genuine efficacy on the subsequent trajectory of creation, while preserving the creator's capacity to reject the system's default direction and redirect the process. The framework consists of four conditions: diachronic reason formation, reflective uptake, trajectory efficacy, and contestability and redirection, together with a minimal tracing requirement. The paper further translates this normative framework into a set of reflective self-interrogation tools for creators working with generative AI.

Ultimately, the paper argues that human authorship under conditions of generative AI depends not on how many steps a person personally performs, but on whether that person's reasons can continuously, reflectively, and effectively shape the development of the work. Dynamic-reflexive tracking therefore constitutes an important normative condition for sustaining creative agency and human authorial attribution, although it is not by itself sufficient to establish authorship in full.
\end{abstract}

\vspace{0.5em}

\begin{quotation}
\end{quotation}

\normalsize

\section{From Creative Outputs to Human Authorship}

Generative AI has become widely integrated into professional workflows across a range of creative fields. Creative practitioners are actively making use of its new capacities while also trying to retain control over AI-assisted creative processes (Tsao et al., 2026). The resulting problem is not simply whether creativity is enhanced or diminished. More fundamentally, generative AI changes how human creative contribution is understood and creates new gaps in the attribution of authorship, credit, and responsibility (Nyholm, 2023).

At the artefact level, the outputs of generative AI have come to satisfy various criteria and definitions traditionally employed in aesthetic theory for judging works of art, thereby emerging as a new creative medium with distinctive affordances (Coeckelbergh, 2023; Epstein et al., 2023). At the same time, the authorship and value of AI art are not determined by the work alone, but are to a significant extent constructed through framing by art markets, platforms, and cultural institutions (van Heerden, 2025). Hence, when creation increasingly relies on generative AI for production, why may a work still be called "someone's" work? If generative AI can produce outputs that are novel, valuable, and aesthetically effective, why does human creation still matter?

To address these questions, we must first clarify the concept of "creativity" adopted in this paper and its evaluative criteria. This paper adopts a comprehensive understanding of creativity. First, creativity involves not only originality and value at the artefact level (Boden, 2004; Gaut, 2010), but also agential contribution in the creative process, namely, a capacity or disposition to produce works that are novel to the creator herself and appropriate in context, rather than merely the fortuitous generation of original and valuable outcomes (Gaut, 2010). Second, creativity is not an attribute of isolated individuals, but a systemic process of dynamic interaction among the individual, the domain, and the field, that is, a continuous interplay between the creator, knowledge resources, and social evaluative mechanisms (Csikszentmihalyi, 2014).

On this understanding, the fact that a work exhibits certain "human-like" creative features does not suffice to demonstrate the presence of a corresponding human creative subject behind it. Emotional expression has traditionally been regarded as a hallmark of human artistic creation (Tolstoy, 1897/1904). However, with the development of generative AI in music, image, and text generation, whether a work expresses or evokes emotion is increasingly inadequate as a sufficient criterion for distinguishing human from AI creation. Experimental studies indicate that listeners are not always able to accurately distinguish between AI-generated and human-composed music; AI-generated music can simulate emotional expression through identifiable musical cues such as rhythm, melody, and vocal delivery, achieving ratings on aesthetic experience and emotional communication comparable to human-created works (Chen  \& Collins, 2026). Thus, emotional expression or affective evocation at the artefact level is no longer sufficient on its own to demonstrate the existence of a sufficient condition for human creative subjectivity behind human-AI collaborative creation.

Concurrently, although the outputs of generative AI exhibit apparent directionality and structure, this "directionality" is not the intrinsic intentionality characteristic of human creative subjects (namely, mental states oriented by the subject's own purposes and accompanied by an understanding of the meaning of one's actions). Rather, it is derived intentionality, manifest through training data, system design, and human feedback, and does not thereby constitute the system as a subject with autonomous purposes and comprehension (Barandiaran  \&  Pérez-Verdugo, 2025). Hence, while AI may generate works that appear to possess intentionality, emotional coherence, and stylistic consistency, these surface features alone do not warrant treating AI as occupying the same creative subject-position as humans.

This apparent subjectivity is not confined to features of the work, but extends to the interactive experience between human and AI. It is therefore necessary to distinguish between AI being perceived as a "collaborative partner" and AI actually possessing legal or ethical authorship and creative agency. In human-AI collaboration, AI may be experienced by users as a teammate or partner; this phenomenon can be understood through the concept of social presence, the degree to which a person, in interaction, perceives another object as having social existence, as if capable of participating as a social agent (Siemon et al., 2026). This suggests that even when AI experientially appears as a collaborative subject, this does not entail that it should be treated as a creative subject in the normative sense.

Therefore, the fact that a work is creative and the fact that a particular person created it as a creative subject are not the same judgement. What is distinctive about Generative AI's entry into creative domains is precisely that it renders these two formerly co-occurring dimensions increasingly separable. This separation occurs not only between an individual creator and a particular work, but also at collective and institutional levels.

\section{Beyond Productive Capacity: Individual, Collective, and Industrial Dimensions of Creative Empowerment}
\subsection{Individual Level: The Expansion of Productive Capacity and the Enhancement of Creative Agency Are Not Necessarily Synchronous
}
Generative AI can significantly lower technical execution thresholds, expand creative search spaces, and increase the speed and volume of content generation (Rafner et al., 2023). It enables individuals without traditional professional training to enter certain creative activities and markets (De Cremer et al., 2023). Through textual prompts, users can translate their intentions more directly into visual, musical, or linguistic outcomes, thereby extending the expressive capacities of non-professional users (Deepa et al., 2025). Those previously constrained by technical training, funding, equipment, or division of labour can now, with generative AI, transform ideas into visible or audible works more rapidly (Doshi  \&  Hauser, 2024). This improvement is particularly pronounced for participants with initially lower creative performance. For instance, individuals without drawing training can generate complex images; those unable to compose can rapidly form musical prototypes; and those lacking film production experience can produce short dramas, animations, or relatively complete audiovisual content.

However, the expansion of productive capacity does not entail a proportional enhancement of human creative agency. The automation afforded by generative systems may reduce some execution burdens, but may also lead creators to offload cognitive activities such as ideation, integration, monitoring, and revision to the system (Rafner et al., 2023). From the perspective of cognitive offloading, while such delegation may improve efficiency, it may reduce the creator's actual cognitive engagement in the work's formation and affect their sense of participation and subjective ownership (Kosmyna et al., 2025). In other words, AI may make it easier for a person to complete a work, yet this does not mean that the person has exercised greater agential contribution over the creative direction during the work's formation.

This decoupling of productive capacity from creative agency may also occur through the system's prediction of past preferences and the shaping of choice spaces, thereby affecting agential contribution. In highly personalised generative systems, the system may increasingly accurately predict the creator's past preferences, yet this does not necessarily enhance the creator's capacity to understand and intervene in the system (Kuilman et al., 2025). Accurate prediction of past preferences is not equivalent to responsiveness to the creator's current creative reasons. A system may know that "I used to like blue," but this does not amount to understanding "why I now have reason to use blue in this particular work." If a personalisation system continuously generates "what users typically like" based on historical behaviour, it may compress the creator's space to deviate from established styles, reassess, and form new directions, thereby affecting the creator's capacity to form and implement the judgments and controls that actually organise the work's development (Atkinson  \&  Barker, 2023).

\subsection{Collective Level: The Enhancement of Model Generative Capacity and the Expansion of Collective Creative Resources Are Not Necessarily Synchronous
}

That generative AI can today "write," "draw," or "compose" does not mean that its exhibited creative capacity is formed independently of human society. Creative activity is always embedded in existing cultural domains, social structures, and evaluative mechanisms (McIntyre, 2022). The knowledge, skills, and expressive forms available to individual creators already incorporate the accumulated outcomes of prior social practice (Marx, 1857–1858). Language, techniques, aesthetic categories, and modes of thought have long been formed through collective human social practice, and the training materials of generative models derive precisely from this accumulated human communication, cultural production, and digital records. AI's exhibited generative capacity therefore remains built upon past and present human collective production, the cultural traces of which—via digitisation—further become conditions for model learning and generation (Halpin, 2026).

Generative AI may thus be understood as an extension, under new technological conditions, of the objectification of social capacities. On the Marxian view, socially accumulated knowledge, skills, and creative resources—accumulated over long periods—may be objectified in machinery and further manifest as the productive power of machinery or fixed capital itself, rather than directly as the capacities possessed by individual labourers. In generative AI, socially accumulated knowledge and creative resources, originally dispersed across human society, are reorganised into generative capacities that models can mobilise at scale.

However, the fact that socially accumulated knowledge and creative resources can be transformed into model capacities does not entail that the growth of model capacities will be re-transformed in the same manner into the growth of human collective creative capacities (Morozov, 2026). First, model generative capacities are primarily sedimented in models and platforms, and access to and use of them are subject to constraints including subscription fees, usage quotas, interface permissions, model ownership, and platform rules. A user's capacity to invoke the model or platform to produce outputs does not equate to the capacity to command this power. Being able to invoke a model for writing, drawing, or composing does not mean one truly masters or disposes of this capacity; once separated from the model or platform, such capacity may not be individually preserved and developed.

Second, the creative experiences, judgments, and solutions formed by individuals through interaction with generative AI do not automatically become collective creative resources. While generative AI lowers creative thresholds and enables more people to acquire the capacity to produce works, this expansion may also be accompanied by the contraction of professional creative labour markets and may undermine the conditions under which some creators can accumulate and transmit high-level creative capacities over the long term (Piskopani et al., 2023). From a collective perspective, genuinely valuable collective intelligence should not assimilate different individuals into a homogenised whole, but should enhance the complementarity of capacities and perspectives among different subjects, thereby sustaining and expanding differentiation (Halpin, 2026).

If creativity is a systemic process of ongoing interaction among the individual, the domain, and the field, then individual practice can be transformed into accumulable and inheritable collective resources only when it is expressed, preserved, and enters the knowledge system of the relevant domain, and is further exchanged, selected, and recognised through creative communities and social evaluative mechanisms (Csikszentmihalyi, 2014). Creative experiences conducted in isolated private AI sessions may vanish with the end of the interaction, whereas models and platforms can continuously accumulate capacities from a large volume of user interactions. Therefore, democratisation in the creative domain cannot be measured solely by how many people acquire the capacity to produce works; one must also examine whether the resulting creative ecology has expanded the differentiation and complementarity of capacities, experiences, perspectives, and expressive modes among different creators.

Finally, generative AI is reshaping how society understands the meaning of creativity and its evaluative standards. Traditionally, artists, works, and technical roles have been gradually constituted through the interplay of humans, technology, and society; their identities depend not only on the work itself, but are also shaped by linguistically and institutionally effective framing practices. Once the art world, the media, and the public widely accept and designate a certain AI-generated work as "art," then whatever objective criteria philosophers may propose, this work is in social practice already treated as art (Coeckelbergh, 2023).

However, this meaning-ascription process is itself shaped by its technological environment. When creators use AI for creation, AI influences and shapes the creator's own judgments and choices through its algorithmic logic and agentic performance (Atkinson  \&  Barker, 2023). Generative systems may also make content increasingly reliant on default model styles, platform recommendation logics, trending templates, and quantifiable market feedback (Chaney et al., 2018). A large volume of highly finished short dramas, short videos, music, images, or texts may be rapidly generated, yet may be highly repetitive and narratively convergent, primarily optimised around click-through rates, conversion rates, and commercial returns. 

Increased productive capacity has not transformed into human creative capacity. On the contrary, the democratisation of generative AI may also lead to the compression of artistic diversity, the devaluation of professional artistic labour, and the monopolistic power of dominant AI companies (Cheng et al., 2024; Doshi  \&  Hauser, 2024). When such technological mediation is extensively and continuously embedded in creation, dissemination, and evaluation practices, its influence is no longer confined to the individual level, but may further shape the shared aesthetic and evaluative referents of social members. 

Consequently, collective understanding and evaluation of "creativity" are often conducted within possibility boundaries partially preset by technological systems. The identification and public evaluation of "creativity" are no longer merely negotiated among the art world, critical communities, and the public, but are increasingly pre-influenced by the design and deployment of generative systems.

\subsection{Industrial Level: The Expansion of Technical Capacity and the Enhancement of Creators' Institutional Power Are Not Necessarily Synchronous
}

This problem further points to the industrial and institutional structure. Generative AI has been adopted by firms as a tool for reducing content production costs and may substitute for certain forms of professional creative labour, thereby compressing creative workers' employment opportunities and bargaining power (Wade, 2024). If such market pressures further undermine the conditions under which professional creators can accumulate and transmit creative capacities over the long term, they may damage the long-term innovative capacity of the creative ecology (De Cremer et al., 2023). This impact is not uniformly distributed across all creative industries: fields more closely tied to copyright regimes and reproducible content production—such as screenwriting and music—are more directly affected by generative AI than creative forms more dependent on liveness, embodiment, or performative context, such as choreography and comedy ("Artificial Intelligence," 2025).

The ongoing controversies in the music industry around "responsible AI" exemplify such institutional conflicts. Major rights holders, technology platforms, creators, and industry organisations exhibit significant disagreements, and these disagreements are underpinned by divergent economic interests and value distributions among different actors. For this reason, "responsible" under the rubric of AI is not an ethical label with a unified meaning, but a normative framework continuously contested and redefined by different actors within concrete industrial relations (Campos Valverde  \&  Kaye, 2026). This indicates that the normative problems of human-AI creation extend beyond who generates the work to include who has the authority to shape the institutions of creation, determine authorial attribution, and participate in the distribution of economic and cultural value.

In sum, the preceding analyses at the individual, collective, and industrial levels indicate that the "empowerment" brought by generative AI is not a unidimensional change. This paper accordingly distinguishes four distinct senses of empowerment:

\begin{itemize}

\item \textbf{Productive empowerment:}
whether AI expands what creators can express and produce;

\item \textbf{Agential empowerment:}
whether creators remain capable of forming, evaluating, and revising
their creative judgments;

\item \textbf{Process autonomy:}
whether creators are able to determine at which stages and in which ways
AI enters their creative labour;

\item \textbf{Distributive justice:}
how the new value co-created by human labour, works, feedback, and data
is distributed among human creators, platforms, and technological actors.

\end{itemize}

These four dimensions are not necessarily synchronous: a system may significantly enhance productive capacity while simultaneously diminishing the creator's judgmental autonomy, process control, or capacity to capture value.

Among these, agential empowerment and process autonomy converge on a more fundamental question of authorship: when a work is co-produced by human and generative system, what kind of human judgment and control suffices for the work to be meaningfully attributable to a human creator?

\section{From Factual Participation to Normative Attribution: Authorship as a Position of Accountability}
The analyses at the individual, collective, and industrial levels jointly indicate that generative AI has pushed the problem in creative domains from "can AI create?" to a more fundamental normative question: when creative agency is distributed among multiple human and machine actors, "to whom creative authorship belongs" becomes itself an open question (Uddin et al., 2025). The application of generative AI in creative domains has extended from mere technical execution to multiple aspects of creative judgment itself. The traditional theory of authorship, which emphasises the human being as the primary source of creative force, faces a complication when AI enters: its "unexpected outputs" may stimulate the creator's inspiration, yet simultaneously blur the boundaries of human judgmental contribution (Kosmyna et al., 2025).

This authorial problem is particularly salient at the legal level. The deep dependence of generative models on existing human works and training data poses new difficulties for copyright attribution, authorship, and ownership of works (Epstein et al., 2023). The legal notion of "author" is not a fixed natural category, but undergoes adjustment with technological, institutional, and social change. In the generative AI context, the determination of authorship cannot be based solely on whether the creator physically or technically executed the work; even when machines directly participate in production, humans may still maintain authorial status through the conception, selection, and realisation of the work. Thus, the conscious choices made by humans in using AI for creation, and the "labour" (Lockean), "personality" (Hegelian), and "communication" (Kantian) they embody, should all be considered in examining the authorship of AI art (Caldwell, 2023).

Clarifying this problem cannot stop at factual descriptions of "who did what," but requires further distinguishing the significance of different contributions for authorial attribution. A person may input numerous prompts, repeatedly generate, and make a final selection, yet operate primarily within the possibilities the system has provided. Conversely, a person may delegate substantial technical execution to AI, yet consistently determine the work's theme, evaluative criteria, structural relations, and creative direction. Neither the volume of manual operations, the proportion of automation, nor the final act of selection can directly determine authorship. What needs explanation is: which human activities merely enable the work to come into existence, and which activities constitute the creative judgments that shape the work?

Answering this question first requires distinguishing creative production capacity from creative judgment. Generative AI can significantly expand creative production capacity, yet this capacity cannot automatically substitute for high-level creative judgment (Saito, 2025). As creative options become increasingly easy to generate in large quantities through AI, the human contribution may instead lie more in evaluation, selection, rejection, integration, and redirection. For instance, in AI-assisted academic writing, a system can fluently organise language, synthesise sources, and propose arguments, yet whether a research question is genuinely important, whether an argument possesses theoretical novelty, and which results merit retention or rejection still require domain knowledge and high-level evaluative judgment (Moorhouse et al., 2025). Generative capacity and creative judgment are therefore not the same: being able to produce an outcome is not the same as being able to judge why that outcome deserves to be part of the work.

Empirical findings on authorship attribution already exhibit sensitivity to this structure of contribution. As the degree of external assistance increases, people's attribution of authorship, creatorship, and responsibility to the human creator all decline, and this change does not depend entirely on whether the helper is human or AI (Formosa et al., 2025). This suggests that authorial attribution is not determined solely by who submits the final work, but shifts with the degree and structure of participation by different actors in the creative process.

Authorial attribution is also influenced by the linguistic framing of AI's role: when AI is described as an "agent" with an "autonomous evaluative framework," the public automatically transfers some credit and responsibility from the artist to the technology developer, even when the artist retains final selection over outputs; the artist's authorship is already diluted in public perception (Epstein et al., 2020). Conversely, if the artist can demonstrate that they control the "evaluative framework" itself, that is, determine what constitutes a good work, which outputs merit retention, and how the work is ultimately integrated, then their authorship is more readily attributable to the human creator rather than to the generative system. This aligns with Rodrigues' (2026) argument that operational control alone is insufficient for authorship; what matters is whether the human dominates the creative process at the higher-order level of judgment, evaluation, and organisation. The creator's reflective endorsement of their creative reasons and evaluative standards, therefore, is not only of philosophical significance but can also be understood as an important practical condition for sustaining authorial attribution.

This fragility of attribution manifests a noteworthy asymmetry in the distribution of credit and blame. Human users are often more readily held accountable for harms caused by AI outputs (since decisions to publish, adopt, and disseminate results are typically still made by humans), yet find it more difficult to claim full credit for AI-assisted outputs, because the public and institutions tend to attribute the "good parts" partly to the system itself. This credit–blame asymmetry (Porsdam Mann et al., 2023) implies that, without establishing a clear normative standard for authorship, human users will systematically bear a structural disadvantage of "losing on both sides", taking full responsibility while sharing credit with the system. This is the practical motivation for this paper's insistence that authorship must be anchored in a human subject capable of bearing normative responsibility. The creative process may be distributed among humans, models, data, and platforms, but authorship still needs to be anchored in a human subject with human answerability. Authorship can be understood as a normative position of "giving and asking for reasons": the creator is able to defend the work, respond to challenges, revise or retract content, and bear corresponding responsibility and consequences. Technical opacity does not negate this accountability requirement; rather, it calls for more precise mechanisms for responsibility tracking and attribution (Uebel  \&  Hamamra, 2026).

\section{From Human Involvement to Meaningful Human Control}

In human–AI systems, high levels of automation can coexist with high levels of human control. Even in complex, context-sensitive tasks involving creative judgment, extensive automation does not by itself imply a loss of human control (Shneiderman, 2020). But this coexistence does not arise automatically. Ethical principles must be incorporated into the design and governance of AI if expanded creative possibilities are to remain compatible with human agency and diversity (Yasuda  \&  Maruyama, 2026). Generative creation therefore requires a more discriminating normative framework: under conditions of high automation, what forms of human involvement count as genuinely meaningful creative control and are sufficient to sustain creative agency and authorship?

Existing ethical and technical literature offers multiple levels of response in human-AI co-creation research. A significant body of HCI literature tends toward operational control mechanisms, focusing on AI's responsiveness to human input; however, such procedural control typically merely describes how humans and AI influence each other (Zhang et al., 2025). This literature addresses how control is implemented and distributed, yet remains insufficient to answer what kind of control can be called "meaningful." Nguyen et al. (2026), in discussions of AI control and governance, distinguish between causal control (whether one can technically intervene in the system) and normative control (whether the intervention is informed, justified, and exercised by a competent authority).

This distinction is particularly clear in the "human-in-the-loop" model: this model emphasises operational intervention capabilities in human-AI co-creative systems (clicking to regenerate, modifying prompts, vetoing outputs), addressing the question of "how humans help machines" rather than whether human participation has substantive significance (Wu et al., 2022). In other words, the fact that a human leaves numerous causal traces in the process does not entail that the system has in a normative sense tracked their creative reasons; creators may, in the course of repeatedly "clicking to regenerate," progressively lose reflexive grasp of their own justificatory grounds. Thus, mere participation or intervention does not suffice to demonstrate that they have retained substantive control at the cognitive and normative levels (Abbink et al., 2024).

Meaningful Human Control (MHC) was proposed precisely to address this limitation as a more demanding evaluative framework. Classical automation and safety-control theory tended toward a "normative responsiveness of the system to human commands, rules, or moral reasons" (Santoni de Sio  \&  van den Hoven, 2018). Subsequently, Pozzi and Santoni de Sio (2026) added that if a system only permits certain model-recognisable preferences to enter the generative process while systematically excluding other experiences, values, or expressions, then formal user participation does not suffice for meaningful human control. The original framework required socio-technical systems to satisfy two conditions: tracking, that the system remains responsive to relevant human reasons; and tracing, that there exist identifiable human subjects who understand the system's capabilities and their own role within it, such that the system's behaviour can be traced back to these human control relations.

MHC is an open and context-dependent normative concept, admitting multiple possible interpretations (Robbins, 2024); its specific requirements depend on the normative purpose it serves, rather than there being a uniform control model applicable to all contexts (Davidovic, 2023). Hence, "what kind of control counts as meaningful" cannot be determined apart from concrete contexts. In the medical domain, for instance, there remains much that is undefined, with principal obstacles including algorithmic opacity, insufficient transparency, and efficiency problems in human-machine collaboration (Hille et al., 2026).

For generative creation, the core purpose of MHC is neither "safety" nor "accountability," but rather sustaining the creator's creative agency and authorial attribution. Therefore, creative attribution cannot be determined simply by whether a human personally executed each step, but must examine whether their judgments and reasons substantively shaped the creative process (Köhler et al., 2025). This is precisely the normative advantage of MHC over merely causal participation-based criteria.

Generative systems need not only to respond to the creator's already expressed instructions, but also to maintain appropriate representation of their continuously evolving creative goals, judgments, and reasons (Cavalcante Siebert et al., 2022), and to allow the creator to contest how the system understands and implements these reasons, with such contestation substantively altering subsequent generation (Calvert, 2025). At the same time, the creator must form a sufficiently accurate understanding of what the system can and cannot do and how it generates outputs. However, the problem is that this capability typically encounters the following situations: users can only "correct" by re-entering natural language, but the system cannot genuinely "understand" why the user is dissatisfied, and its responses to corrections are unstable; a user may reject an output, but this may not change the system's underlying generative bias, and outputs may repeatedly return to similar default styles or patterns; sometimes a user can only adjust within the possibility space the system provides, without being able to reconstitute that space itself.

This distinction can be illustrated with a minimal example. The user inputs "generate a cat in boots," and the system outputs an image matching the description. On the surface, the system appears to have perfectly "tracked" the user's request. However, this form of tracking is extremely impoverished. Formally, the user may have multiple opportunities to operate—repeatedly inputting prompts, adjusting parameters, and selecting from numerous candidates. Yet if the system's default style, model capabilities, and interface structure in fact delimit which creative directions are easily expressible and realisable, then the user's choices may always occur within a highly presupposed possibility space. In such a case, although the user continuously provides input, they may not genuinely allow their creative reasons to shape the work's development. It is not that the system is tracking the user's expressive reasons, but rather that the system's default algorithms and platform aesthetics are surreptitiously tracking, shaping, and narrowing the user's aesthetic preferences.

Personalisation systems further amplify this risk: if a platform discovers through historical data that a user favours a certain style, narrative, or characterisation, and continuously provides the most readily acceptable results, it may possess strong personalisation capability, yet lack strong creative tracking—because what occurs here may not be the development of reasons, but a preference-locking loop:

\textbf{Past preferences}
$\rightarrow$
\textbf{system prediction}
$\rightarrow$
\textbf{similar outputs}
$\rightarrow$
\textbf{user acceptance}
$\rightarrow$
\textbf{additional data}
$\rightarrow$
\textbf{increasingly accurate repetition of past preferences}.

Such a loop may constrain creative emergence by progressively narrowing the space in which new judgments can form. A system that genuinely empowers creators should instead preserve the possibility of a different dynamic:

\textbf{Existing judgments}
$\rightarrow$
\textbf{new possiblities}
$\rightarrow$
\textbf{surprise or conflict}
$\rightarrow$
\textbf{re-evaluation}
$\rightarrow$
\textbf{new reasons}
$\rightarrow$
\textbf{new creative directions}.

Therefore, the most worthwhile goal of generative AI is not to accurately predict the past creator, but to provide the present creator with space to form different judgments—and this is precisely where the original tracking condition is most prone to being misapplied in creative contexts: a system may causally respond perfectly to input, yet normatively never have genuinely tracked the user's reasons.

In resonance with Pozzi and Santoni de Sio's discussion, if a subject is merely treated by the system as an information provider, while their fuller experience, value judgments, and expressive reasons do not actually influence the generative process, then their subject-position may still be diminished. Behaviourally, the human is still involved, but formal participation does not automatically constitute substantive creative control. Participation alone does not guarantee control; the relevant subject must be able to advance their knowledge and reasons in a practically effective manner. Genuinely creative participation should allow the creator to propose, explain, maintain, revise, and even overturn their creative reasons. Therefore, the criterion for creative control must shift from "shallow operations" (input, clicking, selection) to "structural shaping" (whether reasons actually shape the direction of the work's development).

This distinction also has a clarifying implication for the relation between generative AI and non-professional creators: the lowering of technical execution thresholds is not itself the problem. A person may not be able to paint, yet possess very nuanced observations about the life experiences, interpersonal relations, or social phenomena they wish to express; what they lacked was technical execution capacity, not creative judgment. Generative AI can partially decouple the two, undertaking substantial execution work and thereby enabling those previously unable to realise expression to gain opportunities for work formation. In this sense, AI does not necessarily erode creative agency, but may expand it. What truly warrants vigilance is not the pseudo-proposition that "non-professionals' AI creation is inherently inferior," but the distinction between technical execution being outsourced to AI and creative judgment being outsourced to AI. Only the latter concerns the loss of meaningful control that this paper has argued for.

\section{Reconstructing the Tracking Condition}
However, applying the classical MHC tracking condition to generative creation encounters a further, deeper problem: the human reasons that need to be tracked in creative activity are themselves often not given in advance and held constant. In mixed-initiative co-creation and generative AI interaction, creativity is not a static object that is already fully formed at the outset and merely awaits technical realisation. A creator may discover new directions through an unexpected generation, or abandon original goals through repeated comparison, rejection, and revision, coming to a new understanding of what they truly want to express. The human creative intention may thus gradually form, clarify, and change through ongoing interaction with the generative system, rather than existing fully formed prior to interaction and being merely executed by AI (Kreminski \& Chung, 2024). As a new creative medium, generative AI increasingly renders creation a dynamic process from open exploration to intention clarification and work realisation (Epstein et al., 2023).

This means that if tracking is understood merely as the system appropriately responding to a pre-existing human intention or reason, it is insufficient to characterise control relations in generative creation. Basing authorship on "the initial idea came from the human" or "the first prompt was human-proposed" equally overlooks the creator's ongoing shaping of direction, exploration of uncertainty, and revision of intentions in subsequent interactions. What needs to be tracked is not a fixed initial intention, but how the creator's reasons form, change, and continue to influence the work's development over time.

This paper therefore proposes extending tracking in creative MHC to dynamic-reflexive tracking (DRT). DRT requires not only that the generative system respond to the creator's already formed reasons, but also that the human-AI creative process leave space for the formation, testing, and revision of these reasons, and that changes in reasons be capable of effectively reorienting subsequent generation. In other words, whether a work still meaningfully belongs to the creator depends not on whether she remained faithful to an initial intention, but on whether the new reasons that continually emerge during the work's development remain subject to her evaluation, acceptance, rejection, or revision, and exert actual influence on the final creative trajectory.

This process is not a linear movement from a fixed initial reason directly to the final work, but a continuously cycling dynamic process: the creator first enters creation with some preliminary reason; the AI's output then becomes a new object of evaluation; the creator re-evaluates, revises, or even changes the original reason in light of this output, and the new reason further influences the next round of generation. Thus repeated, the final work is not the simple realisation of the initial intention, but a product formed through the ongoing interaction of human reasons, AI outputs, and creator evaluations.

More specifically, dynamic-reflexive tracking comprises at least four interrelated normative conditions.
First, diachronic reason formation: the creator's reasons should be capable of forming and changing during the creative process in response to new materials, unexpected outcomes, and evaluations, rather than being fixed to the initial prompt or past preferences.

Second, reflective uptake: changes in reasons must be capable of being, in some manner, evaluated, accepted, rejected, or re-understood by the creator, rather than merely being passive shaping of preferences by system outputs. This does not require that all reasons can be immediately verbalised; it also includes tacit, perceptual, and experiential judgments.

Third, trajectory efficacy: the formation and change of reasons must be capable of actually altering subsequent choices and the direction of the work, rather than merely being rationalised after the outcome has been produced.

Fourth, contestability and redirection: the creator should be able not only to reject a particular output, but also to have such rejection alter the subsequent process and, when necessary, initiate new directions that the system has not proactively offered.

Together, these four conditions advance the question of creative control from "whether the user performed inputs, clicks, and selections" to a more demanding question: have the creator's reasons continuously participated in and actually shaped the trajectory of the work's development over time?

At the same time, this paper does not thereby abolish the tracing condition in MHC. Even if creative reasons can dynamically shape the work, there still needs to exist an identifiable human subject who possesses task-relative understanding of the AI's specific role, capabilities, and limitations in the creative process, and who can bear corresponding answerability for the key choices in the work. This paper therefore retains a minimal tracing requirement, but does not further reconstruct tracing itself. The reason is that the specific difficulty addressed in this paper is not primarily "whether the responsible subject can be identified," but rather that the creator's reasons themselves form dynamically in human-machine interaction, and how these dynamic reasons can still maintain controlling efficacy over the work. This problem first arises within the tracking condition

\section{Operationalisation: A Set of Self-Interrogative Diagnostic Tools}
If dynamic-reflexive tracking is to function as a normative standard for actual creative practice, creators need some way of examining whether their reasons are still genuinely shaping the work.

The broader analysis in this paper distinguishes four dimensions: productive empowerment; agential empowerment; process autonomy; and distributive justice. These dimensions should not be treated as identical to the four conditions of DRT. Rather, they provide a wider context for asking what kind of empowerment generative AI actually produces.

The creators considered here occupy a position between professional model engineers and completely passive end users. They do not need to understand backpropagation, but they do need to interpret outputs, evaluate them, reject them, and redirect the system. Existing interface and explanatory tools remain comparatively underdeveloped for users of this kind. At the level of control mechanisms, Zhang et al. (2025) identify twelve design strategies, with guided input interaction and iterative feedback loops among the most common. These mechanisms constitute much of the operational toolkit for control in current human–AI co-creation systems. Yet they primarily concern how users operate AI, not how users reflect on and test the reasons underlying their own judgments. This gap motivates a complementary form of control at the level of creator self-reflection.

To make the preceding criteria more usable in practice, the paper borrows a structural feature of venture-capital due diligence: the systematic questioning of founders (Gompers et al., 2016; VC Boom Editorial, 2026). The point is not to import the values of venture capital into creative practice. Rather, founder due diligence often tests whether a decision reflects a genuine understanding of a problem, a market, and the founder’s own position, or whether it merely follows external trends. In particular, it probes why a course of action was chosen, what evidence would change the decision, and under what conditions the founder would abandon or redirect it. This structure closely resembles the reflective and counterfactual questions at the heart of dynamic-reflexive tracking.

\definecolor{headerblue}{HTML}{DCEAF7}
\definecolor{rowblue}{HTML}{F7FBFF}
\definecolor{leftcol}{HTML}{EEF5FB}
\definecolor{dimgray}{HTML}{444444}

\begin{table}[htbp]
\centering
\captionsetup{justification=centering}
\footnotesize
\renewcommand{\arraystretch}{1.28}
\setlength{\tabcolsep}{6pt}

\caption{Reflective self-interrogation tool for creators using generative AI}
\label{tab:reflective-tool}

\begin{tabularx}{\textwidth}{
>{\raggedright\arraybackslash\color{dimgray}}p{0.22\textwidth}
>{\raggedright\arraybackslash}X
>{\raggedright\arraybackslash\color{dimgray}}p{0.18\textwidth}
}
\toprule
\rowcolor{headerblue}

\makecell[l]{\textbf{Question often put to a}\\
\textbf{founder}}
&
\makecell[c]{\\
\textbf{Reflective question for the creator}}
&
\makecell[l]{\textbf{Normative}\\
\textbf{dimension}}
\\

\rowcolor{rowblue}
\cellcolor{leftcol}\textbf{Why this problem?}
&
\textbf{Why am I making this work? If I were not using AI, would I still want to express this issue?}
&
Distal reason
\\

\cellcolor{leftcol}\textbf{Why you?}
&
\textbf{What experience, knowledge, observation, or standpoint gives me a judgment of my own about this subject?}
&
Tracing / standpoint
\\

\rowcolor{rowblue}
\cellcolor{leftcol}\textbf{Why this solution?}
&
\textbf{Why am I keeping this character, structure, image, or melody rather than another one?}
&
Tracking
\\

\cellcolor{leftcol}\textbf{What evidence changed your mind?}
&
\textbf{What did the AI produce that genuinely changed my earlier view, and why did it change my mind?}
&
Dynamic tracking
\\

\rowcolor{rowblue}
\cellcolor{leftcol}\textbf{What would make you reject this plan?}
&
\textbf{What kind of AI output would I reject even if it were visually impressive or technically fluent, and why?}
&
Contestability
\\

\cellcolor{leftcol}\textbf{Can you pivot?}
&
\textbf{Can I formulate a direction the system did not itself offer, rather than merely choosing among A, B, and C?}
&
Redirection
\\

\rowcolor{rowblue}
\cellcolor{leftcol}\textbf{Do you understand the product?}
&
\textbf{Do I know what the AI actually did in this work---which parts it suggested, rewrote, or restructured?}
&
Tracing / understanding
\\

\cellcolor{leftcol}\textbf{Are you just following the market?}
&
\textbf{Does this choice reflect my own judgment, or am I choosing it because the AI or platform tells me that it is what people prefer?}
&
Reason vs.\ preference
\\

\rowcolor{rowblue}
\cellcolor{leftcol}\textbf{What is your moat?}
&
\textbf{If another person entered a similar prompt, would they receive almost the same work? What in this work reflects my own structure of judgment?}
&
Expressive attribution
\\

\cellcolor{leftcol}\textbf{Who controls the key decisions?}
&
\textbf{Were the most important turning points, trade-offs, and stylistic changes decided by me, or produced by system defaults?}
&
Meaningful control
\\

\rowcolor{rowblue}
\cellcolor{leftcol}\textbf{Can you explain your decisions?}
&
\textbf{For the three to five most important choices in the work, can I explain why I chose this rather than that?}
&
Reason-giving
\\

\cellcolor{leftcol}\textbf{Are you learning?}
&
\textbf{Has using AI made me better at judging, comparing, and expressing, or only faster at producing?}
&
Creative empowerment
\\

\rowcolor{rowblue}
\cellcolor{leftcol}\textbf{Who captures the value?}
&
\textbf{Do my labour, work, feedback, and data mainly strengthen my own capacities, or those of the platform?}
&
Distal / systemic control
\\

\cellcolor{leftcol}\textbf{Does the project create durable value?}
&
\textbf{Does the work add a new form of expression, knowledge, or perspective, or mainly reproduce existing templates?}
&
Collective creative value
\\

\bottomrule
\end{tabularx}
\end{table}

\clearpage
This paper's framework has clear boundaries:

First, it does not offer a legal judgment of copyright. Legal authorship depends on the specific jurisdiction, institutional purpose, and configuration of rights, and cannot be directly derived from this table.

Second, it does not offer a complete metaphysical theory of whether AI possesses independent moral agency or should be regarded as a co-author.

Third, it does not claim that all authorship requires the same intensity of explicit reflection; the salient choices differ across media, practices, and collaborative structures.

Fourth, it does not provide an empirical measurement scale. Future research needs to translate indicators such as version history, rejection trajectories, counterfactual redirection, and user understanding into observable design variables.

This table is therefore best understood as offering, for human control in generative creation, an evaluative language with more explanatory power than "did the person click" or "what proportion of AI was used," and as explaining why authorship requires continuity at the level of reasons.

The creative MHC proposed in this paper mainly analyses the individual-level control relationship between the creator and the generative system—that is, whose reasons actually shaped the development of the work? However, even if a creator is able to reject system output, change the direction of creation, and have their own reasons continually influence the work, this does not mean the entire creative ecology possesses an equivalent degree of autonomy. A creator may still find themselves within an institutional environment in which visibility, commercial metrics, and evaluative standards have already been set in advance by the platform. Beyond the individual-level creative MHC (who determines the reasons that shape the work?), there is therefore a related normative question: who determines the goals of the creative ecosystem? This latter question concerns whether the community of creators can participate in forming the goals, values, and directions pursued by creative practice itself.

\section{Main Objections
}

\subsection{Does Dynamic-Reflexive Tracking Demand Too Much in the Way of ``Reasons''?}

Many artistic judgments are intuitive, embodied, or tacit, and creators may be unable to give a complete verbal account of them at the moment of action. Reflective uptake, as used here, therefore does not require creators to articulate a fully explicit set of reasons at every stage.A creator may be able to say only that ``something is wrong here.'' If that judgment nevertheless manifests itself consistently in discrimination, rejection, revision, and redirection, and remains open to later reflection, it can still function as a creatively significant reason. Reason-giving is evidence of reflective control, but it is not its only form.

\subsection{Human Creative Reasons Are Never Formed in Complete Independence}

Artists' judgments are always shaped by education, peers, markets, cultural traditions, and existing media. AI's involvement in reason formation therefore cannot by itself establish a loss of creative agency. The account offered here does not require creative reasons to originate wholly ``from within'' the subject. What matters is whether external influences can be reflectively absorbed, questioned, revised, or rejected.

AI may introduce a new direction or even change a creator's earlier judgment without thereby diminishing agency. Such interaction may instead expand creative agency. The problem arises when system influence becomes increasingly difficult to identify, contest, or redirect. The relevant question is therefore not whether creative reasons are externally influenced, but whether the creator can incorporate those influences into her own deliberative process.

For the same reason, the degree of automation cannot serve as an inverse measure of creative agency. A creator may delegate extensive technical execution to AI while retaining substantive control over themes, structure, evaluative standards, and direction. Conversely, a person may make many prompt revisions and filter numerous outputs while doing little more than react to possibilities supplied by the system. The relevant distinction is therefore between outsourcing technical execution and outsourcing creative judgment. The former may significantly expand creative capacity, especially for people without conventional technical training. What requires scrutiny is the latter when it occurs without corresponding capacities for reflection, evaluation, and redirection.

\subsection{Creative Work Is Often Distributed Across Multiple Agents}

Film, music, design, and many forms of AI-assisted creation are already multi-agent processes, so a work need not be anchored in a single author. The present account does not deny co-authorship or distributed authorship. Dynamic-reflexive tracking does not require every work to be attributed to one independent subject. Its claim is more limited: where a particular share of creative credit is attributed to a person, it should be possible to explain how that person's judgments and reasons actually shaped the relevant creative decisions. The framework is therefore compatible with plural authorship.

\subsection{Creators Can Invent Reasons After the Fact}

The ability to provide a reason after the fact does not show that the reason played any role in the creative process. Faced with a successful AI-generated result, creators may engage in post-hoc rationalisation, retrospectively claiming that ``this is exactly what I intended.'' Reflective endorsement alone is therefore insufficient.

Dynamic-reflexive tracking additionally requires trajectory efficacy. The formation or revision of a reason must make a practical difference to choices, rejections, or subsequent direction during the creative process. If a creator claims to care deeply about a principle, yet changes in that principle make no difference to her behaviour or to the trajectory of the work, the alleged reason is difficult to regard as part of a genuine control relation. The relevant question is not whether a creator can later provide an interpretation of the work, but whether her reasons made a counterfactual difference to how the work developed.

\subsection{Does Criticism of Personalisation Imply That Personalisation Necessarily Narrows Creative Space?}

No. Personalisation may improve a system's ability to respond to users' needs and reduce technical friction, thereby enhancing expression. The objection is not to personalisation as such, but to the possibility of a preference-locking loop, in which systems repeatedly reproduce existing preferences until past behavioural patterns begin to displace the formation of present reasons. The relevant question is therefore not how personalised a system is, but whether it continues to leave genuine room for surprise, contestability, and redirection.

\section{Theoretical and Methodological Limitations
}

First, creative tracking should not be understood as requiring the generative system to possess human-like understanding. The distinction drawn earlier between intrinsic and derived intentionality means that AI need not itself become a minded subject capable of understanding a creator’s reasons. What is required is a functional form of reason-responsiveness: when the creator revises a judgment, objects to an output, or redirects the task, those changes should be capable of producing a reasonably stable and practically effective difference in subsequent generation. Tracking concerns the control relation between human reasons and system behaviour, not the possession of intrinsic understanding by the system.

Second, the four-dimensional account of “empowerment” has a deliberately limited scope. Productive empowerment, agential empowerment, process autonomy, and distributive justice are related but distinct normative dimensions, and they should not all be treated as components of MHC. Creative MHC primarily addresses agential empowerment and process autonomy. Productive empowerment forms part of the wider background of capability expansion, while distributive justice concerns political economy and institutional justice rather than tracking or tracing directly. MHC should therefore not be expanded into a general theory of every social and institutional problem raised by generative AI.

Third, the reflective self-questioning tool is not a strict authorship test. A verbally articulate creator may be very good at explaining her decisions without having exercised genuine control over the creative process. Conversely, a creator who finds it difficult to verbalise tacit judgments may nonetheless possess stable and effective creative agency. The tool should therefore be understood as a reflective diagnostic heuristic for examining whether a creator’s reasons continue to shape the work, not as a mechanical means of proving authorship. Future empirical research would need to compare creators’ self-reports with actual creative trajectories, revision histories, rejection behaviour, and changes of direction.

Fourth, dynamic-reflexive tracking is currently a normative reconstruction rather than an empirically validated measurement model. It asks what kinds of control deserve to count as creatively meaningful; it does not directly measure degrees of creative agency in real-world practice.

The paper also concentrates primarily on end creators using generative AI and does not provide a comprehensive account of the relations among model developers, data contributors, platform designers, and cultural institutions. Creative MHC at the individual level therefore mainly answers the question “whose reasons shape the work?” It cannot by itself answer the broader question “who determines the goals and values of the creative ecosystem?” Addressing that problem requires additional resources from theories of collective autonomy, platform governance, and distributive justice.

\section{Conclusion}

Generative AI makes it increasingly difficult to treat “the work is creative” and “a human being is the creative subject of the work” as the same judgment. AI can generate works with novelty, aesthetic effect, and emotional expression, but these product-level properties do not establish the presence of the same kind of creative subjectivity associated with human creators. The paper has therefore shifted the question from “Can AI create?” to “What makes a work still meaningfully attributable to a human?”

The answer cannot be based on the amount of manual labour, the number of prompts, or the degree of automation. AI may perform extensive technical execution while the human retains creative judgment over themes, evaluative standards, structure, and direction. What matters is not how much the human did, but whether her judgments and reasons actually shaped the work. Productive empowerment through generative AI therefore does not necessarily coincide with agential empowerment, process autonomy, or improvements in distributive position.

MHC provides an important framework for this problem because it moves the analysis of control beyond formal participation towards reason-responsiveness and traceability to relevant human agents. Yet in generative creation, the creator’s reasons are often not fully formed from the beginning. They develop through AI outputs, comparison, rejection, and revision. For this reason, the paper has proposed dynamic-reflexive tracking: what needs to be tracked is not a fixed initial intention, but the way a creator’s reasons form, are evaluated and revised through interaction, and continue to shape the development of the work.

Dynamic-reflexive tracking contains four basic requirements: reasons must be able to form and change over time; those changes must be reflectively taken up or rejected by the creator; changes in reasons must make an actual difference to the trajectory of the work rather than serving merely as retrospective explanations; and the creator must retain the ability to contest system defaults and redirect the process. DRT therefore provides a normative standard for analysing creative agency and human authorship, but it does not by itself provide a complete account of authorship.

The central claim of this paper is consequently a limited but substantive one. Human authorship in the age of generative AI does not depend on how many stages of production a person performs herself. It depends on whether she remains able to form, evaluate, and revise her creative reasons and to make those reasons genuinely alter what the work ultimately becomes. The contribution is therefore twofold: to reconstruct tracking within creative MHC as dynamic-reflexive tracking, and to translate that normative framework into a reflective tool that creators can use in practice. Technical execution can be outsourced, and creative intentions can change. What remains central to meaningful human authorship is the creator’s capacity to evaluate, take ownership of, reject, and redirect those changes.

\end{document}